\documentstyle[11pt,epsfig]{article}

\def\ltap{\raisebox{-.4ex}{\rlap{$\sim$}} \raisebox{.4ex}{$<$}}
\def\gtap{\raisebox{-.4ex}{\rlap{$\sim$}} \raisebox{.4ex}{$>$}}
\newcommand{\TeV}{{\rm\,TeV}}
\newcommand{\GeV}{{\rm\,GeV}}

\newcommand{\eV}{{\rm\,eV}}

\begin{document}
\vspace*{-1in}
\newcommand{\Rslash}{{\not \! \!{R}}}
\renewcommand{\thefootnote}{\fnsymbol{footnote}}
\begin{flushright}
IFUP--TH 67/96\\
\texttt{hep-ph/9611243} \\
\end{flushright}
\vskip 53pt
\begin{center}
{\LARGE{\bf Naturalness constraints on gauge-mediated \\supersymmetry
            breaking models }}
\vskip 30pt
{\bf Gautam Bhattacharyya\footnote{email: gautam@ipifidpt.difi.unipi.it}}
and
{\bf Andrea Romanino\footnote{email: romanino@ibmth.difi.unipi.it}} 
\vskip 10pt
{\it Dipartimento di Fisica, Universit{\`a} di Pisa\\
and INFN, Sezione di Pisa, I-56126 Pisa, Italy }

\vskip 50pt
 
{\bf Abstract}
\end{center}
 
\begin{quotation}
The question of naturalness is addressed in the context of
gauge-mediated supersymmetry breaking models. Requiring that $M_Z$
arises naturally imposes upper limits on the right-handed selectron
mass in these models that are stronger than in the Minimal
Supersymmetric Standard Model and are interesting from the point of
view of searches at the current and future colliders.

\end{quotation}

\setcounter{footnote}{0}
\renewcommand{\thefootnote}{\arabic{footnote}}
\vfill
\clearpage
\setcounter{page}{1}
\pagestyle{plain}

How natural it is to conceive large superpartner scalar masses keeping
the weak scale light? In this Letter we address this question of
naturalness in the context of gauge-mediated supersymmetry breaking
(GMSB) models \cite{PION1,PION2}. There is a naturalness problem in
the Standard Model itself, following from the assumption of the Higgs
boson as a fundamental object, which is best cured by the elegant
introduction of supersymmetry, providing a cut-off for the quadratic
divergences associated with the Higgs.  Since supersymmetry has the
virtue of providing a radiative mechanism for spontaneous breakdown of
electroweak symmmetry, the superpartner scalar masses cannot be
arbitrarily heavy, if we require that the weak scale arises
naturally. In other words, as those scalar masses are pushed higher,
the weak scale still remains light only at the price of an
increasingly delicate cancellation among the soft supersymmetry
breaking masses in the theory. Thus a quantitative requirement of
naturalness puts an upper limit on superparticle masses. In the
Minimal Supersymmetric Standard Model with universal boundary
conditions, this requirement leads to a $\sim$ TeV scale upper bound
on the superpartner masses \cite{BG}\footnote{This issue has been
reinvestigated with a somewhat modified criterion of naturalness in
ref. \cite{AC} and with nonuniversal boundary conditions in
ref. \cite{DG}.}. We find that in GMSB models, the naturalness upper
limits are comparatively stronger.

In the GMSB scenario there exists a set of heavy chiral superfields,
called the ``messenger'' fields.  Supersymmetry is broken in the
messenger sector due to the interaction of the messengers with a
GUT-singlet spurion field which has a non-vanishing vacuum expectation
value (vev) in its auxiliary component ($F$).  The information of
supersymmetry breaking is then transmitted to the superpartners of the
quarks, leptons, gauge and Higgs bosons {\em via} the usual SU(3)$
\times $SU(2)$ \times $U(1) gauge interactions. The gauge-mediation
attributes these models the virtue of automatically suppressing the
Flavor-Changing Neutral Currents.  In the supergravity scenario, where
supersymmetry breaking in the hidden sector is communicated to the
visible sector by gravitational strength interactions, the breaking
scale $\sqrt{F}$ is $\sim 10^{10}$ GeV in order that the scalar masses
obtained from $\tilde{m}^2 \sim F^2/M_{Pl}^2$ are in the TeV range. On
the other hand, in the GMSB models, where the scalar masses are
generated radiatively, the analogous expressions of the squared masses
have a loop-suppression factor of $(\alpha/4\pi)^2$ and $M_{Pl}$ is
replaced by the messenger mass ($10 \TeV \ltap M \ltap M_{Pl}$).
Depending upon the messenger scale, this can lead to a smaller
intrinsic supersymmetry breaking scale and consequently to a
superlight gravitino $\tilde{G}$ constituting the lightest
supersymmetric particle.  Of late, there has been a resurgence of
interests \cite{EEGG} in these models as these massless gravitinos
help to provide a comfortable explanation to the recent CDF event with
$e^+e^-\gamma\gamma~+ $ missing energy in the final state
\cite{CDF}. Very recently, detailed phenomenological analyses of a
general class of GMSB models have been presented in refs. \cite{DTW}
and \cite{BMPZ}.

The simplest version of the messenger models involves a set of
vector-like superfields $(M_i+\overline{M}_i)$. These couple to a
gauge singlet superfield $X$ through the following superpotential,
\begin{equation}
W = \lambda_i X M_i \overline{M}_i. 
\end{equation} 
$M_i$ and $\overline{M}_i$ transform under a complete SU(5)
representation and there could be $n_5$ copies of $(5+\overline{5})$
and $n_{10}$ copies of $(10+\overline{10})$ messengers.  The quantity
$n = (n_5 + 3 n_{10})$ could at most be four from the requirement of
perturbative unification.  The superfield $X$ must acquire non-zero
vevs both in its scalar and auxiliary components through its
interaction with the hidden sector fields.  Assuming for simplicity
$\lambda\equiv \lambda_i$ (for each $i$), it follows from the
superpotential written above that the messenger fermions acquire a
supersymmetric mass\footnote{We denote henceforth the superfields and
their scalar components by the same symbols.} $M \equiv \lambda
\langle X \rangle$ and the messenger scalars are split as having
masses $M_\pm^2 = M^2 \pm \lambda \langle F_X\rangle$. The effective
supersymmetry breaking scale is $\Lambda \equiv \langle
F_X\rangle/\langle X \rangle$.  Integrating out heavy messenger fields
generate gaugino masses at one-loop at the $M$-scale, as
\begin{equation} 
\tilde{M}_i (M) = n~\Rslash~ \tilde{\alpha}_i(M)~ \Lambda
~g\left(\Lambda/M\right),
\label{gaugino} 
\end{equation}  
where\footnote{Our $\alpha_1$ always corresponds to the one which
unifies with $\alpha_2$ and $\alpha_3$, i.e. it is related to the
Standard Model $\alpha_1'$ by $\alpha_1 = 3\alpha_1'/5$.}
$\tilde{\alpha}_i \equiv \alpha_i/4\pi$ correspond to the gauge
couplings and $g(x)$ is the loop function.  $\Rslash$ ($\leq 1$)
arises from the fact the U(1)$_R$-symmetry could break at a scale
smaller than $M$ rendering the gauginos lighter. The scalar masses
arise at the two-loop level and their expressions at the generating
scale $M$ are given by
\begin{equation}
\tilde{m}^2 (M) = 2~n~\Lambda^2 ~f\left(\Lambda/M\right)~\sum_{i=1}^3
C_i ~\tilde{\alpha}_i^2(M),
\label{scalar} 
\end{equation}
where $C_i = 4/3, 3/4$ for the fundamental representations of SU(3),
SU(2) respectively and zero for singlets, $C_i = 3Y^2/5$ for U(1) ($Y
\equiv Q - T_3$) and $f(x)$ is the associated loop function. The
functions $g(x)$ and $f(x)$~\cite{martin} are very close to unity in
the limit $x\rightarrow 0$ and, therefore, the expressions of the
gaugino and scalar masses become particularly simple when $\Lambda \ll
M$, which is a reasonable limit that we rely upon in the rest of the
paper\footnote{The limit $\Lambda = M$ leads to massless messenger
scalars which are definitely unwanted. Also $\Lambda>M$ leads to
unphysical messenger scalar masses.}. Also, in this limit the gaugino
and scalar masses are independent of the Yukawa couplings appearing in
the superpotential. We do not treat $n$ as an independent parameter,
since it can always be absorbed in a redefinition of $\Rslash$ and
$\Lambda$ as $\sqrt{n}\, \Rslash \rightarrow \Rslash$ and $\sqrt{n}\,
\Lambda \rightarrow \Lambda$. Actually, $g$ and $f$ can also be
absorbed in a further redefinition of $\Rslash$ and $\Lambda$.

As has been pointed out in ref. \cite{DvGP}, there is an intrinsic
``$\mu$-problem'' associated with the $\mu \hat{H}_u \hat{H}_d$ term
in the superpotential and the $B \mu H_u H_d$ term in the scalar
potential, since these Peccei-Quinn symmetry-violating terms cannot be
generated by gauge interactions.  Both are generated at the same loop
level in generic models rendering either $\mu$ to be at the weak scale
and $B$ too large beyond the naturalness bound, or $B$ at the weak
scale and $\mu$ too small to be phenomenologically acceptable. The
problem can be cured by arranging the model in such a way that $\mu$
is generated at one-loop but $B\mu$ at two-loop~\cite{DvGP}. This
keeps both $\mu$ and $B$ at the weak scale but indeed at the price of
introducing additional interactions which could affect the Higgs boson
soft mass terms. We parametrize this effect by adding two quantities
$\Box_u^2$ and $\Box_d^2$ at the scale $M$ in the following way:
\begin{equation} 
m_{H_u (H_d)}^2 (M) =  m_{\tilde{l}_L}^2 (M) + \Box_u^2 (\Box_d^2).
\label{box} 
\end{equation} 

Soft trilinear $A$-terms are zero at $M$. Their radiative generation
requires the simultaneous violation of the U(1)$_R$ symmetry and the
chiral flavor symmetry in the observable sector in the broken
supersymmetry phase. Since the messengers do not violate observable
sector chiral flavor, the $A$-terms are not generated at
one-loop. They are generated only at the two-loop level requiring a
gaugino mass insertion in the internal line \cite{DTW}.

The GMSB models that we consider are characterized by the following
parameters: $M$, $\Lambda$, $\Rslash$, $\Box_u^2$, $\Box_d^2$, $\mu$
and $B_0$ (the value of $B$ at $M$).  They determine the electroweak
scale $M_Z$ and $\sin 2\beta$ through
\begin{eqnarray}
\label{ew1}
M_Z^2 & = & 2 \,{{m_{H_d}^2 - m_{H_u}^2\tan^2\beta}
\over{\tan^2\beta - 1}} - 2 \mu^2,  \\
\label{ew2}
\sin 2\beta & = & {{2 B \mu}\over{m_{H_u}^2  + m_{H_d}^2 + 2\mu^2}};  
\end{eqnarray}
where we have used tree-level Higgs potentials.  We compute
$m^2_{H_{u(d)}}$ using the boundary conditions in
eqs.~(\ref{gaugino}), (\ref{scalar}) and (\ref{box}) and running down
from $M$ to the stop mass $m_{\tilde{t}}$ by one-loop renormalization
group (RG) equations\footnote{This is better than running up to $M_Z$
when we consider the tree level Higgs potential.}. This is enough for
the purpose of studying naturalness.  We use $\alpha_3 (M_Z) = 0.117$
and the top and bottom quark running masses at the top mass scale as
165 and 2.8 GeV respectively.

Now we are all set to discuss the fine-tuning criteria. In general, if
a quantity $X \equiv X (a_1, a_2,...)$ is determined as a function of
many input parameters $a_i$, then a measure of how much the parameters
$a_i$ are fine-tuned in the determination of the quantity $X$ can be
expressed through the parameter \cite{BG}
\begin{equation} 
\Delta_{a_i}^X \equiv 
\left| {{a_i}\over X} {{\partial X}\over{\partial a_i}}\right|, 
\label{delta}
\end{equation}
for each $a_i$. $\Delta$ measures the relative variation of $X$
against the relative variation of $a_i$. In our case, a large
fine-tuning (as defined in eq.~(\ref{delta}))
corresponds to large cancellations among independent terms.

Let us first examine the amount of fine-tuning that goes in the
determination of $\sin 2\beta$ through eq.~(\ref{ew2}). For large
$\tan\beta$, $\sin 2\beta$ is small and a cancellation is possible
only if $B = B_0 + B^{\rm RG}$ (where $B^{\rm RG}$ is the running
effect of $B$ and that does not depend on $B_0$) is much smaller than
$B_0$. The quantity $|B^{\rm RG}/B|$, which is in fact the same as
$\Delta^{\sin 2\beta}_{B_0}$, becomes larger if $M$ increases (because
in this case $B^{\rm RG}$ increases) or if $\tan\beta$ becomes larger
(because in this case $B\propto \sin 2 \beta$ decreases). For example,
for $\tan\beta\simeq 50$, $M=10^5\TeV$, $\Delta^{\sin
2\beta}_{B_0}\simeq 10$.

Now let us concentrate on the determination of $M_Z$ through
eq.~(\ref{ew1}). $M_Z$ can be expressed as a function of two sets of
parameters: $\Lambda$, $M$ and $\Rslash$ in one set and $\mu$,
$\Box_u^2$ and $\Box_d^2$, which depend on unknown hidden sector
couplings, in another. Neglecting the running of $\Box^2_{u(d)}$
(which are small), one can rewrite eq.~(\ref{ew1}) as
\begin{equation}
M_Z^2  =  2\,{{\tilde{m}_{H_d}^2 - \tilde{m}_{H_u}^2\tan^2\beta}
\over{\tan^2\beta - 1}} - 2\tilde{\mu}^2, 
\label{ew1prime}
\end{equation} 
where, $\tilde{m}^2_{H_{u(d)}} = m^2_{H_{u(d)}} (\Box^2_u = \Box^2_d =
0)$ and
\begin{equation} 
\tilde{\mu}^2 = \mu^2 - {{\Box_d^2 -
\Box_u^2\tan^2\beta}\over{\tan^2\beta - 1}}.
\label{mutilde}
\end{equation}
From eq.~(\ref{mutilde}), it is clear that all the hidden sector
couplings could effectively be expressed just in terms of one
parameter through a redefinition of $\mu^2 \rightarrow \tilde{\mu}^2$
and, therefore, there is no loss of generality even if we set
$\Box^2_u = \Box^2_d = 0$ in our subsequent analysis.

Fine-tuning, in our case, corresponds to large cancellations between
the two {\em uncorrelated} quantities in the right hand side (RHS) of
eq.~(\ref{ew1prime}). A quantitative measure of the cancellation
between these two terms is given by
\begin{equation} 
\Delta \equiv \left|{{2\mu^2}\over{M_Z^2}}\right| = 
\left|2\,{{m_{H_d}^2 - m_{H_u}^2\tan^2\beta}\over{(\tan^2\beta - 1)
M_Z^2}} - 1\right| = \Delta^{M_Z}_\mu.
\label{deltamz}
\end{equation}
Owing to our ignorance of the hidden sector couplings,
$\Delta=\Delta^{M_Z}_\mu$ is the best available measure of fine-tuning
in this set-up.
The rest of our paper concerns a qualitative and a quantitative
analysis of $\Delta$. 

First, we observe that for large $\tan\beta$, $\Delta \simeq
2|m_{H_u}^2/M_Z^2|$. In this case, the squared top Yukawa coupling
($Y_t$) is insensitive to $\tan\beta$, while the bottom- and
tau-Yukawa couplings do not contribute to $m^2_{H_u}$ renormalization.
Therefore, for fixed $\Lambda$ and for $\tan\beta \geq 6$, $\Delta$
does not vary with $\tan\beta$. Actually, for large $\tan\beta$,
$m_{\tilde{\tau}_R} \ltap m_{\tilde{e}_R}$ owing to nonnegligible
$Y_{\tau}$-radiative corrections, so that, if we fix
$m_{\tilde{\tau}_R}$ instead of $\Lambda$, a small dependence of
$\Delta$ on $\tan\beta$ creeps in. In any case, we hereafter confine
ourselves to moderate $\tan\beta$ region.

Apart from $\tan\beta$, $\Delta$ depends also on $M$, $\Rslash$ and
$\Lambda$ (or $m_{\tilde{e}_R}$). Before demonstrating their exact
quantitative effects, let us examine their qualitative dependences by
varying them one at a time while keeping the others fixed. From
eqs.~(\ref{gaugino}) and (\ref{scalar}) and the RG evolution it
follows:
\begin{eqnarray}
\label{hd}
m_{H_d}^2 & \simeq & {3\over2}\left[\tilde{\alpha}_2^2(M) + \Rslash^2
\left(\tilde{\alpha}_2^2(M) - \tilde{\alpha}_2^2(m_{\tilde{t}})\right)\right]
\Lambda^2, \\
\label{hu}
m_{H_u}^2 & \simeq &m_{H_d}^2 - 3\int_0^{t(m_{\tilde{t}})} dt~ \tilde{Y}_t 
     (m_{\tilde{Q}_3}^2 + m_{\tilde{t}_R}^2 + m_{H_u}^2),   
\end{eqnarray} 
where, $t(Q) = 2\ln(M/Q)$ and $\tilde{Y_t}=Y_t/4\pi$. In the RHS of
eq.~(\ref{hu}) the second term is roughly proportional to
$\alpha_3^2$, and, neglecting the Yukawa radiative corrections (just
for qualitative understanding), the dependences on $\alpha_3^2$ itself
turn out to be
\begin{equation}
m_{\tilde{Q}_3,\tilde{t}_R}^2 \approx 
\left[{8\over3}\tilde{\alpha}^2_3(M) + {8\over
9}\Rslash^2\left(\tilde{\alpha}_3^2(m_{\tilde{t}}) -
\tilde{\alpha}_3^2(M)\right)\right] \Lambda^2. 
\end{equation} 
From eq.~(\ref{deltamz}) it follows, 
\begin{equation} 
\Delta = \left({{\Lambda}\over{M_Z}}\right)^2 H(M,\Rslash,\tan\beta) -
1,
\label{deltafinal}
\end{equation} 
where, actually, $H$ has a mild logarithmic dependence on $\Lambda$
through $m_{\tilde{t}}$, as is apparent from
\begin{eqnarray}
\label{H}
H & \approx & {{6 \tan^2\beta}\over{\tan^2\beta -1}}
\int_0^{t(m_{\tilde{t}})} dt~ \tilde{Y}_t\left[{16\over
3}\tilde{\alpha}_3^2(M) + {16\over 9} \Rslash^2
\left(\tilde{\alpha}_3^2 - \tilde{\alpha}_3^2(M)\right)\right]
\nonumber \\ & & -3 \left[\tilde{\alpha}_2^2(M) +
\Rslash^2\left(\tilde{\alpha}_2^2(M) -
\tilde{\alpha}_2^2(m_{\tilde{t}})\right)\right].
\end{eqnarray}
Owing to the $\alpha_3^2$-dependence, the first term on the RHS of
eq.~(\ref{H}) is the leading one (also, this term is responsible for
electroweak symmetry breaking) and, consequently, $H$ is positive. The
prefactor $(\Lambda/M_Z)^2$ in eq.~(\ref{deltafinal}) can be expressed
in terms of the right-handed selectron mass as
\begin{equation} 
\left({{\Lambda}\over{M_Z}}\right)^2 \simeq
{{5\over{6\tilde{\alpha}_1^2(M)}}}
\left[{{m^2_{\tilde{e}_R}}\over{M_Z^2}} - \sin^2\theta_W
{{\tan^2\beta-1}\over{\tan^2\beta+1}}\right],
\label{selectron}
\end{equation}
where, for illustration, we do not explicitly show the small U(1) RG
correction.  In the above equation, the term containing
$\sin^2\theta_W$ corresponds to the tree-level U(1) $D$-term
contribution.  For given values of $\tan\beta$, $M$ and $\Rslash$, $H$
is determined and this could be translated into an upper bound of
right-handed selectron mass. For example, for $\tan\beta = 6$, $M =
100$ TeV and $\Rslash = 1$, $m_{\tilde{e}_R} <$ 83 (280)$\GeV$ for
$\Delta <$ 10 (100) (these numbers follow from exact numerical
analysis). We emphasize that for a light selectron, the tree-level
$D$-term relaxes the constraint. We also notice that the experimental
lower limit on $m_{\tilde{e}_R} >$ 45 (70) GeV from LEP1 (LEP1.5) can
be translated through eq.~(\ref{selectron}) to a lower limit of
$\Lambda$ depending on $M$; for example, for $M\sim \Lambda$, $\Lambda
\gtap$ 10 (30) TeV and, on the other extreme, for $M\sim M_{\rm GUT}$,
$\Lambda \gtap$ 4 (15) $\TeV$.

In the moderate $\tan\beta$ region and for given values of $M$,
$m_{\tilde{e}_R}$ and $\Rslash$, a lower $\tan\beta$ increases
$\Delta$ implying a larger fine-tuning and hence a stronger
constraint. Decreasing $\tan\beta$ increases $\Delta$ because the
leading term in $H$ contains a factor $\tan^2\beta/(\tan^2\beta-1)$,
$Y_t(m_t) \propto (\tan^2\beta+1)/\tan^2\beta$ and finally for a given
$m_{\tilde{e}_R}$, $(\Lambda/M_Z)^2$ increases (see
eq.~(\ref{selectron})).

Other parameters remaining fixed, if the U(1)$_R$ symmetry breaks at a
scale lighter than $M$, $\Rslash < 1$ in the minimal GMSB model ($n_5
=1$, $n_{10} = 0$), implying a lower $\Delta$ and hence a weaker
constraint. If U(1)$_R$ breaks at $M$, then having more messengers
increases $\Rslash$ (from 1 to at most 2 remaining consistent with the
requirement of perturbative unification) and the constraint becomes
stronger.

The dependence of $\Delta$ on $M$ is not apparent unlike in the cases
of other parameters owing to several counteracting
contributions. $\Delta$ depends in fact on $M$ through the range of
integration, the boundary conditions on gaugino and scalar masses and
the $\alpha_1^{-2}(M)$ dependence of $\Delta$ in
eq.~(\ref{deltafinal}) in conjunction with eq.~(\ref{selectron}). It
turns out that $\Delta$ decreases when $M\rightarrow M_{\rm GUT}$.

Now we concentrate on exact numerical analysis.  The dependences of
$\Delta$ on $M$ and $\Rslash$ are plotted in Fig.~1a while in Fig.~1b
we exhibit the limits on $m_{\tilde{e}_R}$ as functions of $\tan\beta$
for conservative choices of $M$ and $\Rslash$ extracted from
Fig.~1a. Combining eqs.~(\ref{deltafinal}) and (\ref{selectron})
yields
\begin{equation}
\Delta \simeq \left({{m^2_{\tilde{e}_R}}\over{M_Z^2}} - \sin^2\theta_W 
            {{\tan^2\beta - 1}\over{\tan^2\beta + 1}}\right)
            \frac{5}{6\tilde{\alpha}_1^2(M)} H(M,\Rslash,\tan\beta)-1
\end{equation}
\begin{figure}[tbhp]
  \begin{center}
\epsfig{file=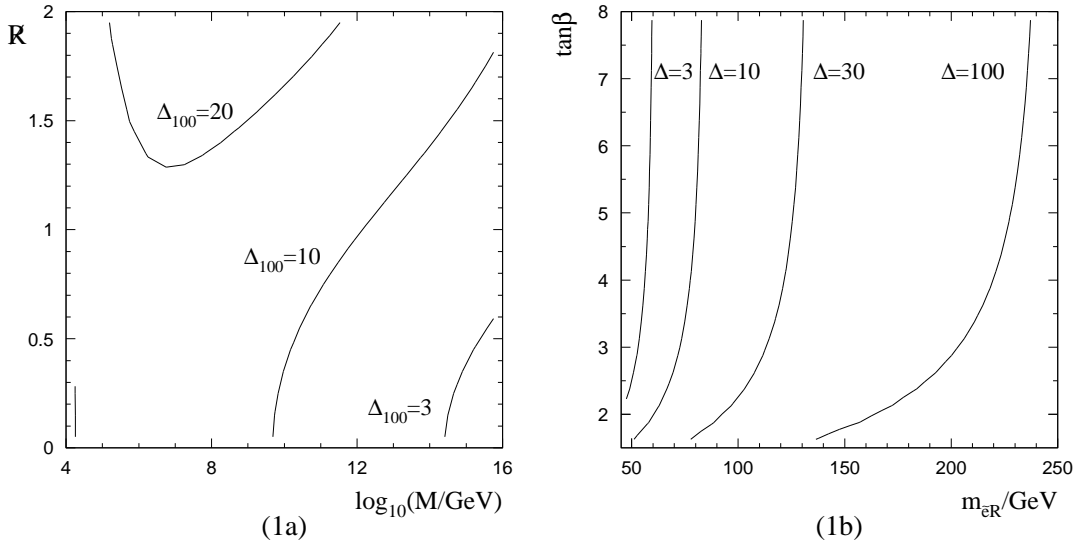,width=\textwidth}
\caption{\em (a): Contour plots of $\Delta_{100}$ (the fine-tuning
parameter for $m_{\tilde{e}_R} = 100\GeV$) in the
$\log_{10}(M/{\rm GeV})$--$\Rslash$ plane for $\tan\beta = 6$; ~(b)
naturalness upper limits on the right-handed selectron mass 
at fixed $M = 10^5\TeV$ and $\Rslash = 1$.}
\label{GMSB}
   \end{center}
\end{figure}
For studying $\Rslash$ and $M$ dependences, we fix $\tan\beta =
6$. First we choose a representative value $m_{\tilde{e}_R} = 100$ GeV
and denote the fine-tuning parameter as $\Delta_{100}$. Then we plot
isocurves of $\Delta_{100} (= 3, 10, 20)$ in the $\log_{10}(M/{\rm
GeV})$--$\Rslash$ plane. The fine-tuning for an arbitrary value of
$m_{\tilde{e}_R}$ can be read from Fig.~1a, using
\begin{equation}
\Delta \simeq \left({{m^2_{\tilde{e}_R}}\over{M_Z^2}} - 0.21\right)
\Delta_{100},
\end{equation} 
which is a good approximation unless $m_{\tilde{e}_R}$ is in a
physically uninteresting region, being much heavier than 100 GeV.
It is apparent from Fig.~1a that in the region $M\sim M_{\rm GUT}$ a
right-handed selectron as heavy as $100\GeV$ does
not suffer from fine-tuning problems. Also lowering $\Rslash$ reduces
the fine-tuning. 

In Fig.~1b, we show the naturalness upper limit on $m_{\tilde{e}_R}$
as a function of $\tan\beta$ in the moderate $\tan\beta$
regime\footnote{As observed before, $\Delta$ is flat with respect to
$\tan\beta$ for $\tan\beta > 6$.} for $M = 10^5\TeV$ (which
corresponds to $m_{\tilde{G}}\sim 100 \eV$ for
$\Lambda=10\TeV$~\cite{BMPZ}) and $\Rslash =1$. 
   
To conclude, we have studied the naturalness constraints in the GMSB
models. The requirement of a natural determination of $M_Z$ in terms
of the parameters of the model yields rather strong upper limits on
the mass of the right-handed selectron. For example, for $M = 10^5$
TeV, $\Rslash = 1$ and $\tan\beta = 6$, $m_{\tilde{e}_R} <$ 80 (230)
GeV for $\Delta =$ 10 (100). If we choose $\Delta < 10$ (i.e. a
fine-tuning not larger than one order of magnitude), the upper limits
are quite interesting for LEP2. We note in passing that an explanation
of the CDF $e^+e^-\gamma\gamma~+ $ missing energy event requires a
cancellation of one order of magnitude in a large range of $M$. The
quoted upper limits are more stringent than in the Minimal
Supersymmetric Standard Model.  In fact, in the latter scenario, the
universal boundary conditions on the scalar masses do not permit the
lightest scalar to be much lighter than the (up-type) Higgs mass. On
the other hand, in the GMSB case, owing to the proportionality of the
scalar masses to different gauge couplings (squared) at $M$, the
lightest scalar becomes light enough compared to the Higgs mass to
pose a strong fine-tuning constraint.

We thank Riccardo Barbieri for many illuminating discussions regarding
the GMSB models and for reading the manuscript.

After completing our work, we have become aware of a related work 
in ref.~\cite{strumia}.

\end{document}